# Multiparameter admittance spectroscopy for investigating defects in MoS$_2$ thin film MOSFETs

Eros Reato[1,2], Ardeshir Esteki[1], Benny Ku[1,2], Zhenxing Wang[2], Michael Heuken[3,4], Max C. Lemme[1,2] and Olof Engström[2]

[1]Chair of Electronic Devices, RWTH Aachen University, Otto-Blumenthal-Strasse 25, 52074 Aachen, Germany
[2]AMO GmbH, Otto-Blumenthal-Strasse 25, 52074 Aachen, Germany
[3]Compound Semiconductor Technology, RWTH Aachen University, Sommerfeldstrasse 18, 52074 Aachen, Germany
[4]AIXTRON SE, Dornkaulstrasse 2, 52134 Herzogenrath, Germany

Email: lemme@amo.de, Phone: +49 241 8867 200, FAX: +49 241 8867 560

*Abstract—* **A method for assessing the quality of electronic material properties of thin-film metal oxide semiconductor field-effect transistors (MOSFETs) is presented. By investigating samples with MOCVD-grown MoS$_2$ channels exposed to atmospheric conditions, the existence of electron traps in MoS$_2$ and at the interface between the gate insulator and the thin-film MoS$_2$ are revealed. Differential conductance and capacitance data of the transistor channels are plotted as 3D surfaces on a base plane spanned by the measurement frequency versus the gate voltage. The existence of defects is confirmed by comparison with ideal results from a theoretical model.**

## I. Introduction

Industry and academia are looking closely at transition metal dichalcogenides (TMDCs), a class of atomically thin, layered two-dimensional (2D) materials, as potential field-effect transistor (FET) channel materials to reach the ultimate scaling limit of silicon technology [1], [2] and for developing flexible electronics [3], [4]. Among semiconducting 2D materials, molybdenum disulfide ($MoS_2$) stands out as a semiconducting 2D material with a reasonable bandgap of 1.8 eV in its single-layer form [5], making it a great candidate for FET channels[6] and integrated circuits [7], [8], [9]. $MoS_2$ growth techniques are still in the early stages of research, with the goal of large-scale, high-quality growth processes [3]. Device fabrication processes are also immature, as current industrial processes and materials have been optimized mainly for silicon technology. In particular, there is limited research addressing the analysis and characterization of the charge traps, which is still a hot topic in the community.[10], [11], [12], [13], [14], [15], [16], [17]. Previous studies on high-k dielectrics on $MoS_2$ and in classical bulk devices show that these materials are prone to be defective and show a sizable density of oxide traps at specific energies, which interact with the flow of charge carriers in the semiconductor channel[18], [19], [20], [21], [22], [23]. Among the published work, standard techniques on pseudo-MOSFET structures with global back gates have been used to investigate $MoS_2$ interfaces[17], [24], one study analyzed the response of high-k and h-BN as gate dielectrics on an exfoliated $MoS_2$ natural crystal[16] , and a recent attempt has been made with a monolayer $MoS_2$ using the Deep Level Transient Spectroscopy (DLTS) technique [25].

In this work, we investigate the dynamic properties of back-gated FETs fabricated with MOCVD $MoS_2$ films by studying conductance and capacitance data as 3D patterns on a plane spanned by the gate voltage and signal frequency. This technique, named Multi-Parameter Admittance Spectroscopy (MPAS), was earlier developed to characterize thermal emission and capture qualities of electron states at semiconductor/oxide interfaces [19], [23], [26]. Here, we develop a corresponding treatment for investigating material properties of thin-film MoS$_2$ MOSFETs, creating portraits (MPAS-maps) of electron state densities at the semiconductor/gate oxide interface, and identifying defects in the MoS$_2$ film.

**II. Sample preparation and geometry**

The layout of the investigated devices consists of a large area, back-gated FET with a gate length of L = 75 µm and a width of W = 80 µm (**Figure 1a**). Silicon with 90 nm thick thermal SiO$_2$ was used as the substrate for the experiment. First, a Ti metal back gate was deposited on the oxide using lift-off. Second, a 45 nm thick layer of Al$_2$O$_3$ deposited by an OXFORD INSTRUMENTS Plasma-Enhanced Atomic Layer Deposition (PEALD) process at 300°C was used as the gate dielectric, followed by a layer transfer of 2D MoS$_2$ grown by Metal-Organic Vapor-Phase Epitaxy (MOVPE) in an AIXTRON Planetary Reactor on sapphire (0001) substrates [27]. The material is multilayer and the thickness is around 5 nm [8]. We used negative photolithography to expose the contact area. After opening the contacts window, we etched the exposed MoS$_2$ with CF$_4$ and O$_2$ gases in an OXFORD INSTRUMENTS Reactive Ion Etching (RIE) tool and sputtered a 50 nm thick layer of Ni right afterward, followed by a standard lift-off process. This process step allowed us to create source and drain edge contacts to the MoS$_2$ layer [28]. Finally, we defined the channel area by photolithography and RIE etching with CF$_4$ and O$_2$ gases. A cross section schematic of the device structure is shown in **Figure 1a** and a top view of the device can be seen in **Figure 1b.** This layout is expected to create Schottky barriers at the interface between the Ni electrodes and the MoS$_2$, as shown in the schematic in **Figure 1c**.

The measurements were performed with an impedance analyzer (Agilent 4294A) with a frequency range between 100 Hz and 1 MHz, a gate voltage range between -5 V and 5 V, and a small signal amplitude of 25 mV. We probed the devices in atmospheric conditions directly on the substrate with a Cascade Probe station.

### III. Structural properties of the MoS₂ layer

We first carried out Atomic Force Microscopy (AFM) and Kelvin Probe Force Microscopy (KPFM) measurements to get insights into the morphology of the surface and the charge population of the as-grown MoS$_2$. The single scan height measurement of the MoS$_2$ surface in Error! Reference source not found.**a** and the height map shown in Error! Reference source not found.**b,** obtained with AFM, demonstrate that the thickness of the material varies significantly along the surface since the deposited MoS$_2$ is multi-layered and polycrystalline [27], [29]. From KPFM measurements illustrated in Error! Reference source not found.**c**, we notice that the potential fluctuations between the grain boundaries follow the change in the number of layers in Error! Reference source not found.**a** and **b**. In particular, the potential relative to the tip increases with an increasing number of stacked layers. At the boundaries between the layers, charge carrier traps are expected and were found also in the experimental data, later discussed in connection with **Figure 6**.

### IV. Circuit geometry and theoretical considerations

The MoS$_2$ FETs were connected in a circuit geometry as shown in **Figure 3a**. The source and drain contacts are both connected to ground. The gate voltage is driven by a small signal AC-voltage source in series with a DC voltage supply and instrumentation for current and phase determination, all included in the impedance analyzer unit. The Nickel (Ni) source and drain contacts are connected to the end surfaces of the MoS$_2$ channel, establishing back-to-back Schottky structures, as discussed in more detail below in connection with **Figure 3b**. The sinus-shaped small signal, *ΔV$_G$ sin(ωt)*, where *ΔV$_G$* is the amplitude, *ω* is the angular frequency, and *t* is time, creates a displacement current across the gate oxide, driving an AC-current from both ends of the channel, as discussed in detail below in

**Figure 3c**. This AC voltage gives rise to a small energy shift of the Fermi-level position of the channel, thus filling and emptying electron states within a limited, differential energy range of the MoS$_2$-conduction band, playing the role of an AC current pump.

**Figure 3b** shows the shape of the MoS$_2$ conduction band edge (red color) as a function of distance from one of the two S/D contacts at the channel ends, where the Ni-contacts give rise to Schottky structures (compare also **Figure 1c**). Electrons in the metal, with an energy distribution sketched by the black curve starting at the metal Fermi-level, $E_{FM}$, will be able to pass into the MoS$_2$ conduction band at energies above the red conduction band maximum. This rounded shape occurs by mirroring the negative electron charge in the metal when passing between the metal and the semiconductor [30]. The gate oxide, shown in blue, is imagined behind the semiconductor and is, in turn, covered by the gate metal (see **Figure 1a**). As long as the applied gate voltage is $\Delta V_G = V_G = 0$, the whole structure is at thermal equilibrium such that the metal and the semiconductor electrons are governed by the same Fermi-level, $E_{FM}$. This voltage will deviate from zero due to the existence of fixed gate oxide charges and the difference in work function values between the gate and S/D metals (as will be discussed below).

Changing $V_G$ to a positive value will create positive charge at the metal/oxide interface of the gate, which will require a corresponding negative charge in the semiconductor conduction band, delivered from ground and passing the Schottky barrier. That will result in an excess carrier concentration in the channel such that thermal equilibrium expires in that part of the structure. Those electrons are injected along the channel, to equilibrate among the energy band states, but not with the semiconductor lattice. Their energy distribution and concentration are defined by a quasi-Fermi-level $E_{FS}$, as noted in orange in **Figure 3b**. Since both Source and Drain (S/D) contacts are connected with ground, no electric field occurs along the channel. The shape of the conduction band edge is, therefore, constant, except for the depletion regions of the Schottky barriers. The excess electron concentration is kept in place by the perpendicular electric field between the gate metal and the channel. Similarly, since the S/D metals cover the MoS$_2$ end surfaces, the barrier height is only affected by an electric field along the channel. As a result, $E_{FS}$ is lowered within the depletion layer of the barrier region until it coincides with $E_{FM}$ at the point of the length scale where the Schottky barrier has its maximum, as marked in **Figure 3b**. Therefore, at this point, the metal and the semiconductor are both at thermal equilibrium. This means that the carrier concentration at the maximum point of the barrier is kept at the value given by thermal equilibrium as governed by $E_{FS} = E_{FM}$. For an ideal Schottky barrier, this means that the maximum rate of electrons injected into the channel per unit time, $e_{nb}$, (i.e. the current) is that of an unbiased Schottky diode in both current directions and expressed by

$$e_{nb} = q^{-1} A_M A^* T^2 \exp\left[-\frac{\Phi_B}{kT}\right], \qquad (1)$$

Where $q$ is the elementary charge, $A_M$ is the cross-section area of the MoS$_2$ – layer, $A^*$ is the Richardson constant of MoS$_2$, $T$ is the temperature, $k$ is the Boltzmann constant, and $\Phi_B$ is the energy of the Schottky-barrier maximum. This motivates the existence of two diodes arranged back-to-back at the source and drain ends of the channel in **Figure 3a.**

Due to the thickness variation of 1 – 7 monolayers across the gate area (see **Figure 2a**), the small signal voltage amplitude of $\Delta V_G$ = 25 mV gives rise to a variation, $\Delta E_{FS}$, of the MoS$_2$ quasi-Fermi-level in the layer of 3 – 9%. This leads to a corresponding amplitude variation of $\Delta E_{FS}$ in the range of 0.8 – 2.4 meV, meaning that the influence of $\Delta E_{FS}$ is small. Based on Boltzmann statistics, the differential influence $\Delta E_{FS}$ on electron concentration, $\Delta n$, in the conduction band is

$$\Delta n = \frac{N_C}{kT} \exp\left(-\frac{E_C - E_{FS}}{kT}\right) \Delta E_{FS}. \qquad (2)$$

This gives rise to a differential channel conductance, $G_t$, from which the rate of electrons transferred along the channel, can be expressed as

$$e_{nt} = \frac{G_t}{C_{ox}} = \frac{q\mu\Delta n}{C_{ox}} \frac{A_G}{L}, \qquad (3)$$

where $q$ is the electron charge, $\mu$ is the electron mobility of the channel, $C_{ox}$ is the oxide capacitance, $A_G$ is the gate area, and $L$ is the channel length.

The current, represented by the two electron emission rates, $e_{nb}$ and $e_{nt}$, will pass the Schottky barrier and the channel in a series configuration, thus building up an effective electron rate, $e_{neff}$, in the same manner as two conductor elements connected in series, and give rise to

$$e_{neff} = \frac{e_{nb}e_{nt}}{e_{nb}+e_{nt}} \qquad (4)$$

Using this in Eq.(3), we get obtain the differential conductance in response to a small signal voltage step

$$G = C_{ox} \, e_{neff}. \qquad (5)$$

Applying a positive DC gate voltage, $V_G$ will move the quasi-Fermi-level $E_{FS}$ closer to the MoS$_2$ conduction band, which momentarily gives rise to empty energy band states below $E_{FS}$. This opens the possibility for electrons to diffuse into the channel from the S/D contacts and fill up the empty states occurring in the conduction band. A new quasi-equilibrium situation occurs in the conduction band when the concentration of electrons in the channel has increased. If the change in $V_G$ is a step voltage, Eq.(5) suggests that $e_{neff}$ = $G/C_{ox}$ is the reciprocal time constant for carrier injection, which would decrease as an exponential function, exp(-$e_{neff}$ t).

However, the driving voltage, $\Delta V_G$, used for the measurement is a sinus wave, which creates a situation as illustrated in **Figure 3c**. The AC small-signal voltage $\Delta V_G \sin(\omega t)$, applied on the gate, will give rise to a cosinus-shaped displacement current perpendicular to the gate surface, opening up unoccupied electron states in the channel conduction band as described above. This current is schematically shown by the red curve in **Figure 3c**. With a +90 degrees delay, current from the channel S/D contacts start to be injected into the channel, as illustrated by the blue curve. The amplitudes of these two functions depend on the extent of non-occupied states, with concentrations expected to decrease with time. From the reasoning above, it is rational to assume this development to be exponential, which determines the decreasing amplitudes of the current waves passing the channel and the displacement current, as shown by the black curve in **Figure 3c**. Conductance and capacitance values can be extracted from the time integrals, measured over one period of time *1/ω*, of the products between the black exponential curve and the two current curves (red and blue).

The corresponding small signal conductance is expected to have a time and frequency dependence determined by a sinus function. Considering the process recorded over *1/ω*, and replacing the constant $C_{ox}$ with a frequency-dependent capacitance, we get

$$\frac{G(t,\omega)}{\omega} = C(t,\omega) e_{neff} \sin(\omega t) \exp(-e_{neff} t), \tag{6}$$

where the corresponding expression for capacitance would be

$$C(t,\omega) = C_{ox} e_{neff} \cos(\omega t) \exp(-e_{neff} t). \tag{7}$$

The frequency dependence can be calculated with Eq. (6) by integrating along time:

$$\frac{G(\omega)}{\omega} = C(\omega) e_{neff} \int_0^\infty \sin(\omega t) \exp(-e_{neff} t) \, dt = C(\omega) \frac{e_{neff} \omega}{\omega^2 + e_{neff}^2}, \tag{8}$$

where the integral is the Laplace-transform, $\omega/(\omega^2 + e_{neff}^2)$, for sinus. Similarly, for the frequency dependence of the capacitance, we get

$$C(\omega) = C_{ox}\, e_{neff} \int_0^\infty \cos(\omega t) \exp(-e_{neff} t)\, dt = C_{ox} \frac{e_{neff}^2}{\omega^2 + e_{neff}^2}, \qquad (9)$$

where the integral is the Laplace-transform, $e_{neff}/(\omega^2 + e_{neff}^2)$, for cosinus. The final expressions for $G(\omega)/\omega$ and $C(\omega)$ in Eqs. (8) and (9), respectively, describe the dynamic properties of MoS$_2$ MOSFETs as long as we assume the materials assembling the device to be intrinsic.

**V. MPAS - maps**

MPAS-maps of calculated $G(\omega)/\omega$ and $C(\omega)$ data as a function of $V_G$ and log(1/ω) are shown in **Figure 4a** and **b**, respectively. Two ridges are observed in **Figure 4a**, connected with a pronounced kink at about $V_G = V_{G0}$ = 0.55 V. The two electron rates, $e_{nt}$ and $e_{nb}$, combine into the effective value, $e_{neff}$, as given by Eq.(4), where the smallest of the two dominates. The sloping linear ridge on the $V_G$ vs. log (1/ω) plane occurs because of the exponential dependence of $e_{nt}$ on the quasi-Fermi-level position, $E_{FS}$, as expressed by Eqs. (2) and (3). The maximum is a result of resonance attributed to $e_{nt} = \omega$ as given by Eq.(8). The ridge going in parallel with the $V_G$ axis is determined by the constant $e_{nb}$ value defined in Eq.(1). The resonance condition, for this part is $e_{nb} = \omega$. In Fig. 3(b), we find $C(V_G, \omega)$ as a plateau limited inside a $V_G$ vs. log(1/$\omega$) area.

A. MPAS simulation

To create MPAS-maps, i.e. to plot the quantities $G/\omega$ and $C$ on a $V_G$ vs. log(1/$\omega$) plane, one needs to find a relation between $E_{FS}$ and $V_G$. From Eqs.(2) and (3), we compare the ratios $E_{FS}/(\log(e_{nt})$ = 59.5 meV/decade [19], with the corresponding value, $V_G/\log(e_{nt})$ from the experimental slope. Further, adjusting for the gate oxide charge and difference in work function values between Ni and Ti, thus fitting the $V_G$ values by a parallel shift along the measured scale, we find for the sample discussed below

$$E_C - E_{FS} = 0.046\,(6.85 - V_G), \qquad (10)$$

were $E_C - E_{FS}$ has the unit eV. A plot of Eq. (10) is shown in **Figure 4c**. Starting at $V_G$ = -5 V. where $E_{FS}$ is positioned 0.55 eV below the MoS$_2$ conduction band edge, decreasing down to values close to 0.1 eV, meaning that $E_{FS}$ would come very close to the conduction band when approaching $V_G$ = +5V. This would give very high values for $e_{nt}$. However, $e_{neff}$ is impeded by the limited electron rate, $e_{nb}$, passing the Schottky barriers (Eq. (1)). In the theoretical MPAS map in **Figure 4a**, one notices that a kink occurs in the ridge for $V_G$ = 0.55 V. At that point, the electron rate transfers from being limited by $e_{nt}$ to $e_{nb}$, which is determined by the Schottky barrier. For a continued increase of $V_G$, $e_{nb}$ becomes the smaller of the two and takes over $e_{neff}$ as expressed by Eq. (4) and shown in **Figure 4d**. The value used for $e_{nb}$ in **Figure 4** is 2.7 x 10$^5$ [1/s] which corresponds to a Richardson constant of 100 A/cm$^2$ and a barrier height 0.65 eV for T = 300 K in Eq. (1).

The calculated results in **Figure 4a** and **b** are obtained from Eqs. (8) and (9) and thus valid for an ideal sample without defect electron states neither at the oxide/MoS$_2$ interface nor within the MoS$_2$ semiconductor film. The resonance peak heights of the conductance data are equal for both electron rates, $e_{nt}$ and $e_{nb}$, to $C(\omega)/2$ and, in turn, $C(\omega) = C_{ox}/2$, we notice that $G(\omega) = C_{ox}/4$ at the resonance ridges as found from Eqs. (8) and (9). Hence, the deviation of experimental data from the theoretical results in **Figure 4** can be attributed to defects within the investigated sample.

The theoretical MPAS-map for $C(V_G, \omega)$ is shown in **Figure 4b**. From this graph, one finds that for low $\omega$–values the capacitance approaches $C_{ox}$. For common types of measurements of the gate capacitance as a function of gate voltage, with $\omega$ as a parameter, the ledges of the *C-V* curves would move in parallel towards lower gate voltage with lower saturated capacitance values for higher frequency.

**Figure 4c** clarifies the relations between quasi-Fermi-level position $E_C - E_{FS}$ and applied gate voltage $V_G$. For negative $V_G$, the quasi-Fermi-level goes deep into the MoS$_2$ bandgap, such that the concentration of electrons in the conduction band is very small, as quantified by the Boltzmann expression. For a $V_G$ value of +0.55 V, a kink occurs in the MPAS map of **Figure 4a**, where the sloping resonance ridge, regulated by $e_{nt}$, breaks to become the horizontal ridge controlled by the rate, $e_{nb}$, of electrons injected from the Schottky barriers at the channel ends. In **Figure 4c**, this point is marked by the connection between the $e_{nt}$ and the $e_{nb}$ branches. The transition occurs at the point where the two rates are equal. As pointed out above, in relation with **Figure 3b**, the electron rate $e_{nb}$ is ideally expressed by Eq. (1) and determined by the Schottky barrier at thermal equilibrium. Therefore, at the point where $e_{nt} = e_{nb}$, one also has $E_{FM} = E_{FS}$ such that the whole sample goes to thermal equilibrium. This point has the same significance for the present sample geometry as the flat-band voltage ($V_{FB}$) of a three-dimensional MOS structure.

### B. MPAS experiment

**Figure 5a** and **b** display experimental results for comparison with the theory of Eqs. (8) and (9) as plotted in **Figure 4**. The main features of the theoretical MPAS maps are cognizable. The two resonance features, the $e_{nt}$ ridge and the horizontally appearing $e_{nb}$ puckers in the $G/\omega$ - and $C$ – maps can be noted in the theoretical results of **Figure 4** as well as in the experimental **Figure 5**. An additional feature of the experimental data is the gauze-like features in blue to green colors for the slightly leaning plane limited by -7 < log(1/$\omega$)□<□–5 and $V_G$ > 0 existing in both **Figure 5a** and **b** and marked "$D_{it}$"**.** This effect relates to the variation of channel current delivered from the Schottky-barriers at the channel ends, creating small variations in $E_{FS}$ around the equilibrium position $E_{FM}$. This in turn gives rise to electron capture and emission traffic between channel and electron states at the interface between $MoS_2$ and the gate oxide. Such processes are commonly frequency dependent due to limited emission and capture rates [31]. The time constants of such processes differ from those of the rest of the system. Therefore, $G/\omega$ and $C$ do not resonate with the channel electrons, but decrease steadily with increasing frequency, as can be seen in **Figure 5 a** and **b** [32]**.**

From the experimental parallel conductance and capacitance results, one can extract the magnitude and phase angle of the impedance [33]. For the areas shown by white color in **Figure 5a**, we found that their phase angles differed by 180 degrees as compared with areas of positive conductance values. That can be understood by considering a sketch of the $MoS_2$ bandgap as shown in **Figure 6a**. The figure shows the conduction band edge at energy $E_C$, the position of an electron trap at $E_T$, and two positions of the Fermi-level, $E_{FS} + \Delta E_{FS}$ or $E_{FS} - \Delta E_{FS}$, representing the turning points produced by the small-signal amplitude $\Delta V_G$. We pointed out in connection with **Figure 3b** that increasing $V_G$ moves $E_{FS}$ closer to the conduction band edge. This means that, starting with $E_{FS}$ on a trap level at $E_T$, the positive half-period of $\Delta V_G$ will raise $E_{FS}$ by an energy $\Delta E_{FS}$ above $E_T$. Simultaneously, electrons are injected into the channel, whereof a portion will be captured into the trap, thus decreasing the carrier concentration in the channel and the measured channel conductance. For the negative half period of $E_F$, the opposite course occurs. Consequently, the carrier concentration and, thus, the channel conductance is influenced by electron capture into traps with a phase angle of 180 degrees compared with the channel current. Therefore, if the trap concentration in the channel is high, the influence of the trap on the differential channel conductance, $G$, is negative. By plotting one of these white negative-conductance-areas marked as "Neg. G" as $-G/\omega$ on the MPAS map, where the surrounding area is close to zero, we find that its top value is about 0.7 As/V as shown in **Figure 6b**. This value represents a resonance occurring at about $V_G$ = -4.5 V and at a frequency of $\omega$ = 5 kHz. From **Figure 4c**, showing the relation between $V_G$ and $E_C - E_{FS}$, we find that this gate voltage corresponds to $E_C - E_{FS}$ = 0.5 eV. Therefore, it is reasonable to assume that the negative differential conductance demonstrated in **Figure 6b** is the result of a trap complex positioned in the $MoS_2$-layer with a series of energy levels cantered at about 0.5 eV from the conduction band edge

If such a trap in the MoS$_2$ layer occurs with similar resonance data as that of the ridges in the conductance map, one would expect that the amplitude at resonance of the conductance data would be lowered. Such an effect can be seen in **Figure 5a** where the white area exists on both sides of the $e_{nt}$ ridge. This would explain the "$e_{nt}$-valley" marked on the MPAS map. Also, a lowering of the capacitance amplitudes in **Figure 5b** can be noted in areas of the MPAS map, that occur white in **Figure 5a.** The differential channel conductance, G(ω), depends on the charge transported as an ohmic current along the channel with a small cross section area between the two end contacts. The presence of traps decreases the channel charge or even makes its time dependence shift by 180° as illustrated by **Figure 6b**, On the other hand, the capacitance depends on displacement current passing across the MOS-structure with the much larger gate area, where the contribution from charge carrier traffic in trap states has a relatively smaller influence. This explains the modest influence of traps on capacitance in **Figure 5b** as compared with the more dramatic appearance in **Figure 5a** and **Figure 6b**.

**VI. Discussion**

In the modeling of the Schottky barriers at the channel ends of the transistor, we have simply assumed that the electron rate across the barrier maximum for thermal equilibrium is ideal, as described by Eq. (1). This means that we have adopted a hypothesis that the junction at the MoS$_2$/Ni interface is a perfect side contact and flat. In such geometry, the metal protects the interface from the electric field directed parallel with the interface plane such that perturbation of the barrier shape is avoided. However, we established in **Figure 2** that the channel material is polycrystalline, which means that microscopically varying barrier values are highly probable. Therefore, a slight influence of the gate voltage on the barrier height would be expected for $V_G$ values where the electron rate, $e_{nb}$, is limited by injection across the Schottky diode. Consequently, the ridge for positive $V_G$ – values in **Figure 5** and the "$e_{nb}$ – branch" sketched in **Figure 4c** are not perfectly parallel with the $V_G$-axis but have a slight slope towards higher $\omega$ - values and decreasing values of $E_C - E_{FS}$, respectively.

The value of $V_{G0}$, where the kink in the MPAS-map for $G/\omega$ appears, was declared as the point where the two rates of electrons passing the Schottky barrier and those passing along the channel are equal, meaning $e_{nb} = e_{nt}$. The two rates are connected in series, where the smaller value has the strongest influence on the final effective rate, $e_{neff}$, as given by Eq. (4). This means that, at the kink voltage, $V_{G0}$, $e_{neff} = e_{nb}/2 = e_{nt}/2$, (considering Eqs. (1) – (3) and **Figure. 4c** and changing the gate voltage, $V_G$, from low to high values), the quasi-Fermi-level, $E_{FS}$ moves from a position deep in the MoS$_2$ band gap to a position close to the conduction band edge. Therefore, $e_{nt}$ is much smaller than $e_{nb}$ and defines $e_{neff}$. Since $e_{nb}$ is weakly dependent on $V_G$, this rate takes over when $E_{FS}$ passes the kink-point, where $V_G = V_{G0}$. At that point, the whole sample is at thermal equilibrium and $E_{FS} = E_{FM}$ as discussed in relation to **Figure 3c**. For higher $V_G$, where $E_{FS}$ approaches the conduction band, $e_{nt} >> e_{nb}$ and the latter defines $e_{neff}$.

An earlier development of MPAS for silicon MOSFETs focused on thermal electron emission and capture at MOS interface states [19]. Here, the treatment was concentrated on the semiconductor/oxide interface states under the argument that the movement of the Fermi-level at that interface was much smaller than $kT$. This meant that the rate for changing the occupation of a state was close to twice that used in the present analysis, which motivated the factors of 2 times the emission rate in the expressions for thermal electron emission [19]. In our analysis of the present samples, that condition was found invalid for electron transport rates along the transistor channel, which motivates the absence of that factor in Eqs. (8) and (9) above, in agreement with original developments of conductance and capacitance expressions as suggested in Ref. [19].

**VII. Conclusion**

In summary, this study investigates the dynamic properties of field-effect transistors with MoS$_2$ channels fabricated with MOCVD films in a back-gate configuration. We used Multi-Parameter Admittance Spectroscopy (MPAS), a technique originally developed to characterize thermal emission and trapping properties at semiconductor/oxide interfaces in silicon MOSFETs [19]. MPAS creates three-dimensional maps when conductance and capacitance data are plotted as a function of gate voltage and signal frequency. By extending MPAS to analyze thin film MOSFETs, the study aimed to explain MPAS maps of electron state densities at the semiconductor/gate oxide interface and thus identify defects and non-idealities in the MoS$_2$ film.

An analysis of the structural properties revealed significant thickness variations in our MoS$_2$ films, affecting the charge population and morphology. Theoretical considerations, supported by circuit geometry discussions, elucidated the interplay between gate voltage, electron injection, and carrier concentration in the channel. These discussions laid the groundwork for understanding the dynamics of conductance and capacitance in the experimental setup.

The MPAS maps revealed resonance features and provided insight into carrier transport mechanisms. Theoretical predictions agreed with experimental observations, validating the effectiveness of the approach. In addition, the study identified intriguing phenomena, such as negative resonance peaks and gauze-like features, attributed to electron capture and emission processes at the MoS$_2$/gate oxide interface.

Our investigation provides insight into the complex behavior of MoS$_2$-based FETs and suggests a new approach for the analysis of their admittance spectra for the discovery of defects and non-idealities. By integrating theoretical frameworks and experimental validations, this work offers new insights into the characterization of 2D semiconductor/high-k systems and could be further extended to analyze the behavior of other 2D and thin film materials.

## VIII. Acknowledgments

We acknowledge funding by European Union's Horizon 2020 research and innovation program under the grant agreements Graphene Flagship (881603), the German Research Foundation (DFG) through research grants MOSTFLEX (407080863), ULTIMOS2 (LE 2440/8-1) and the DFG Major Research Instrumentation Programme (INST 221/96-1). The authors thank Prof. Andrei Vescan, Dr. Holger Kalisch and Dr. Annika Grundmann from the Chair of Compound Semiconductor Technology (CST) at the RWTH Aachen University for providing the $MoS_2$ layers and for the fruitful discussion.

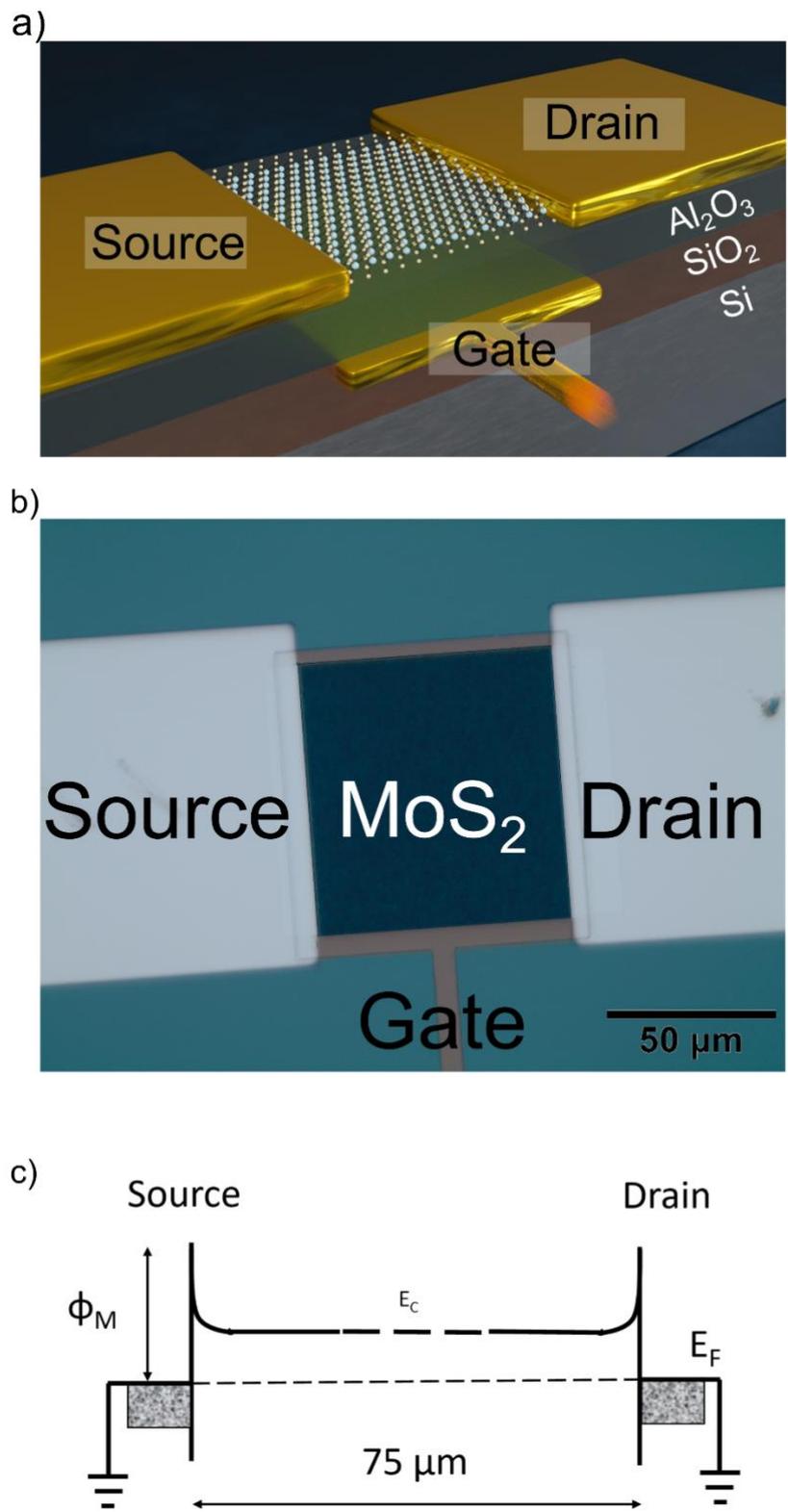

Figure 1 a) Schematic cross-section of the MoS$_2$ device. The gate length is 75 µm and the gate width is 80 µm. b) Optical micrograph (top view) of the MoS$_2$ device. c) Schematic band diagram of the device with the source and drain contacts connected to ground.

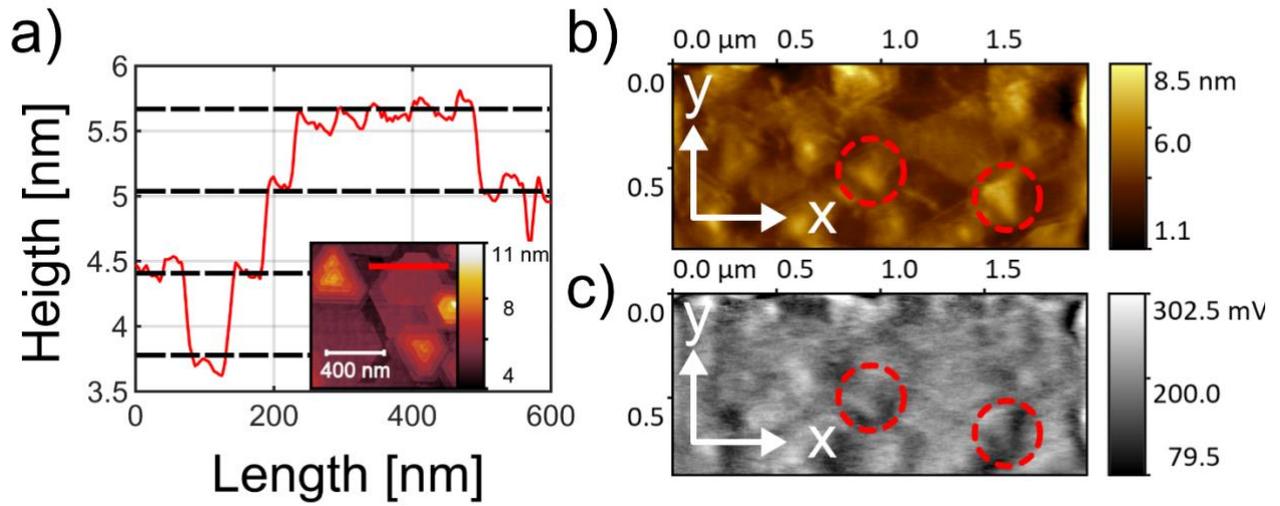

Figure 2 a) AFM single scan of the MoS$_2$ on the growth substrate. Here we could recognize the multilayer nature of the material, with the height of the terraces being around 0.7 nm. b) AFM and c) KPFM scan of a portion of the material. The height measurement shows the terraces while the KPFM shows that the surface potential follows the morphology of the material.

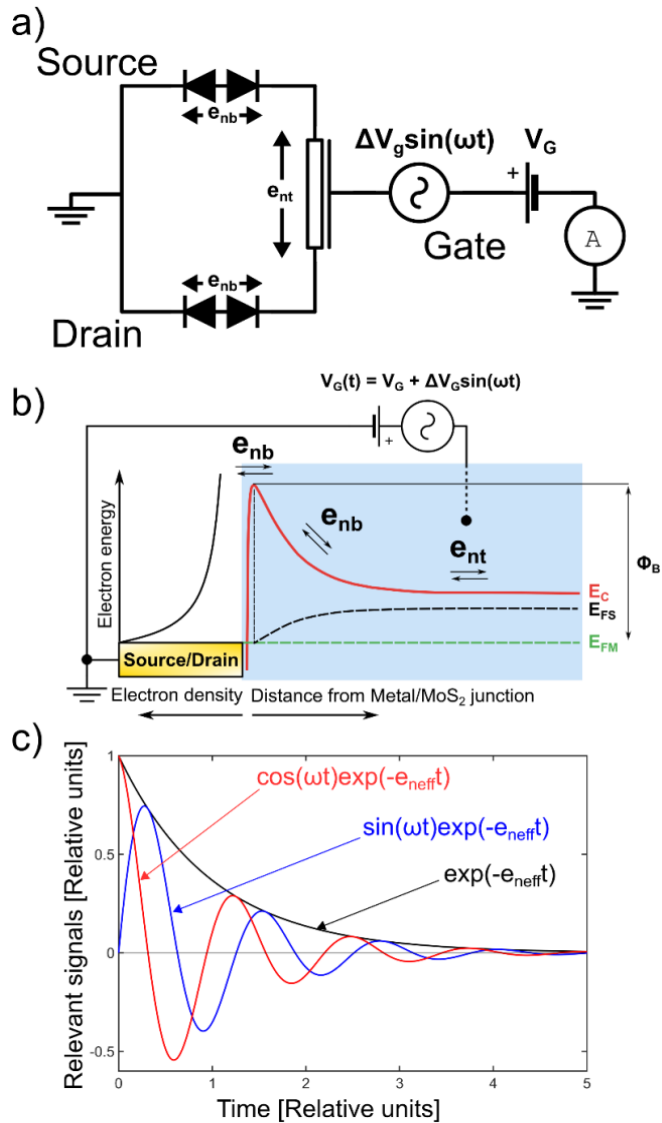

Figure 3 a) Equivalent circuit schematic of the device under test. The signal generator, the DC supply and the meter are included in the impedance analyzer. The gate of the device is connected to the "high" terminal, while the source and drain are connected to the "low" terminal of the Agilent 4294A. b) Schematic of the band diagram of the $MoS_2$ close to the source and Drain terminal. The blue rectangle represents the gate stack. The gate voltage is connected to the back side (see black dashed connector line) The red curve represents the conduction band edge, the green dashed line the Fermi-Level of the metal, while the black curve is the quasi-Fermi level of the channel, a consequence of the electric field applied from the back gate. c)Plot of the currents passing the sample. The red curve is the displacement current passing the gate oxide, opening the channel electron states for the resistive current of the blue curve, injected through source and drain contacts and delayed by 90 degrees. The black curve represents the concentration of empty channel states during one measurement period.

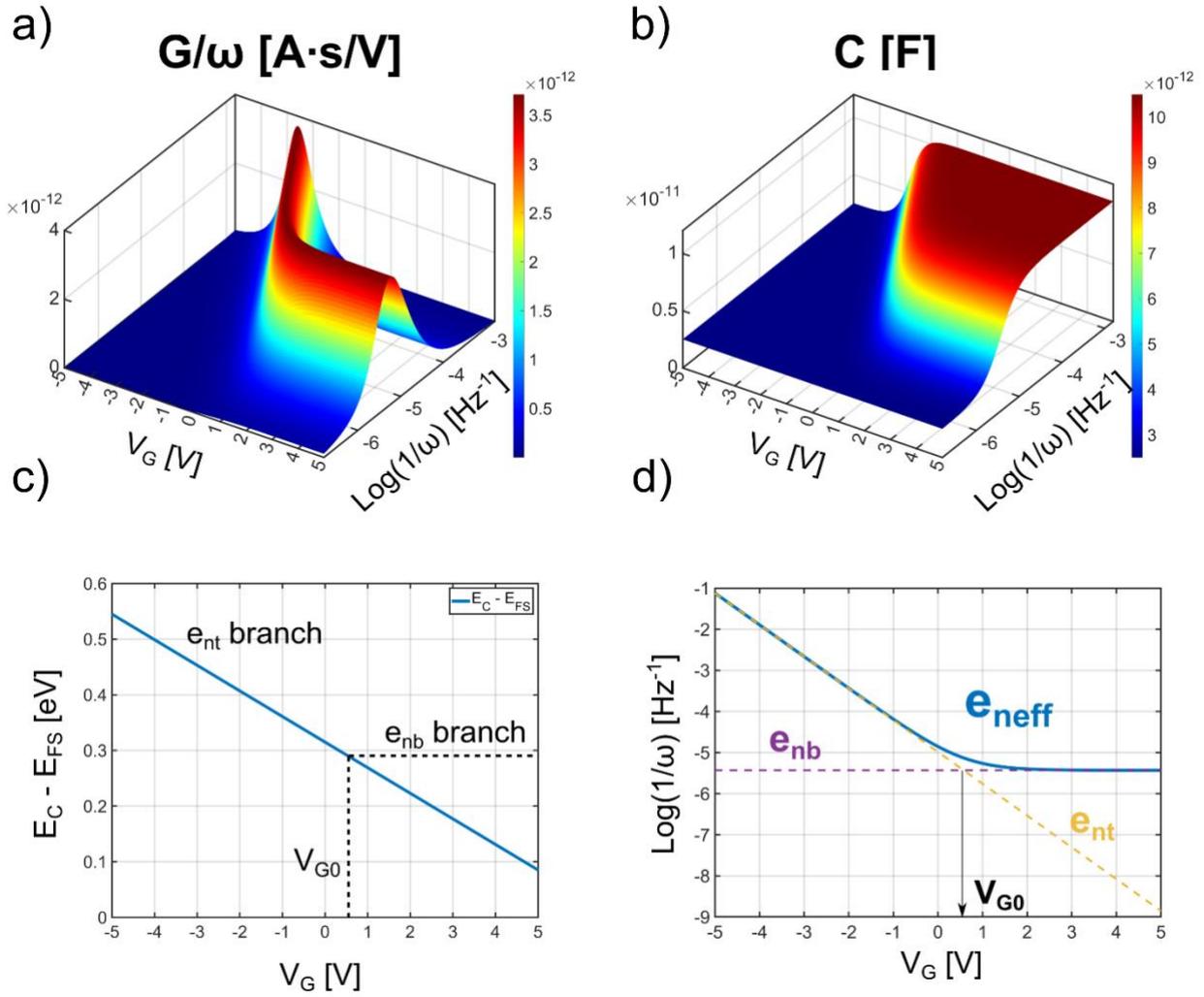

Figure 4. Theoretical MPAS maps of a) conductance, $G/\omega$ and b) capacitance, $C$, both plotted on a $V_G$ vs. Log $[1/\omega]$ plane, where $\omega$ is the angular frequency of the small-signal voltage $\Delta V_G$. Data used for the effective density of states and mobility for electrons were $20 \times 10^{19}$ cm$^{-3}$ [34] and 0.6 cm$^2$/V·s [6], respectively. c) The quasi-Fermi-level position, $\Delta E_{FS}$, as a function of gate voltage $V_G$. d) Plot of the calculated $e_{neff}$ as used for the maps in a) and b). The dashed lines represent the contributions from $e_{nt}$ and $e_{nb}$. The crossing point between the two was used to extract the $V_{G0}$ point at $V_G$ = 0.55 V.

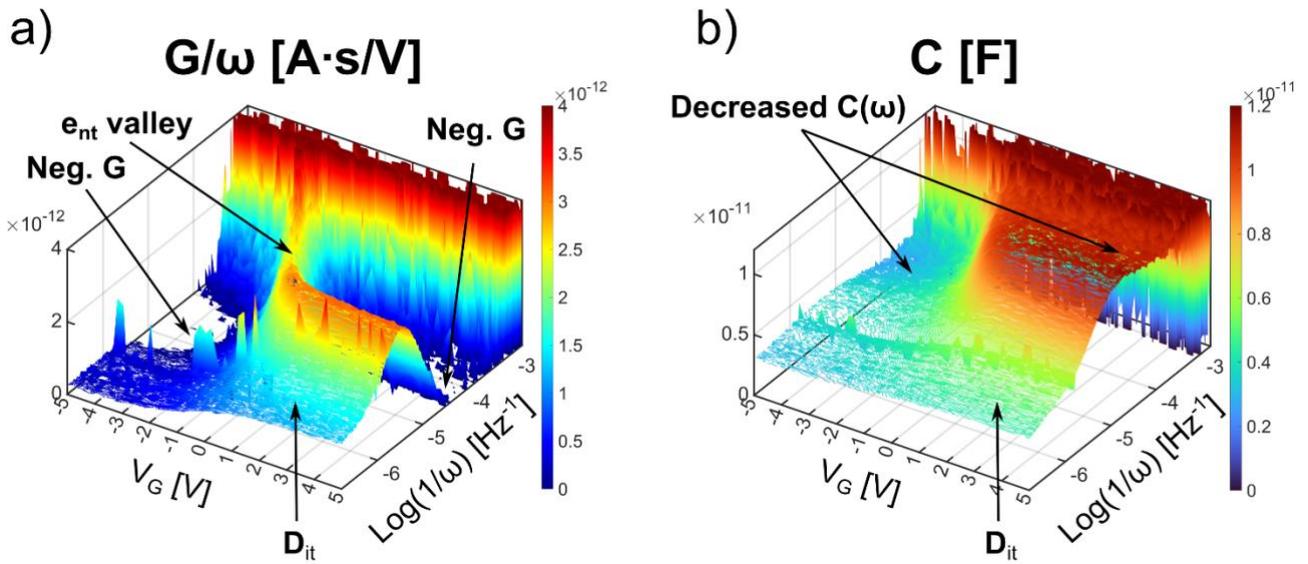

Figure 5. 3D representation of the experimental a) conductance and b) capacitance. The measurement at frequencies below Log(1/ω [Hz$^{-1}$]) = -3.5 (measurement frequencies between 100 Hz and 500 Hz) and the line Log(1/ω [Hz$^{-1}$]) = -5.78 (a measurement frequency of 100 KHz) are resulting from the noise of the instrument and are visible as a the wall-like feature on the backside of the figure and the noisy line in the middle of the figure, respectively.

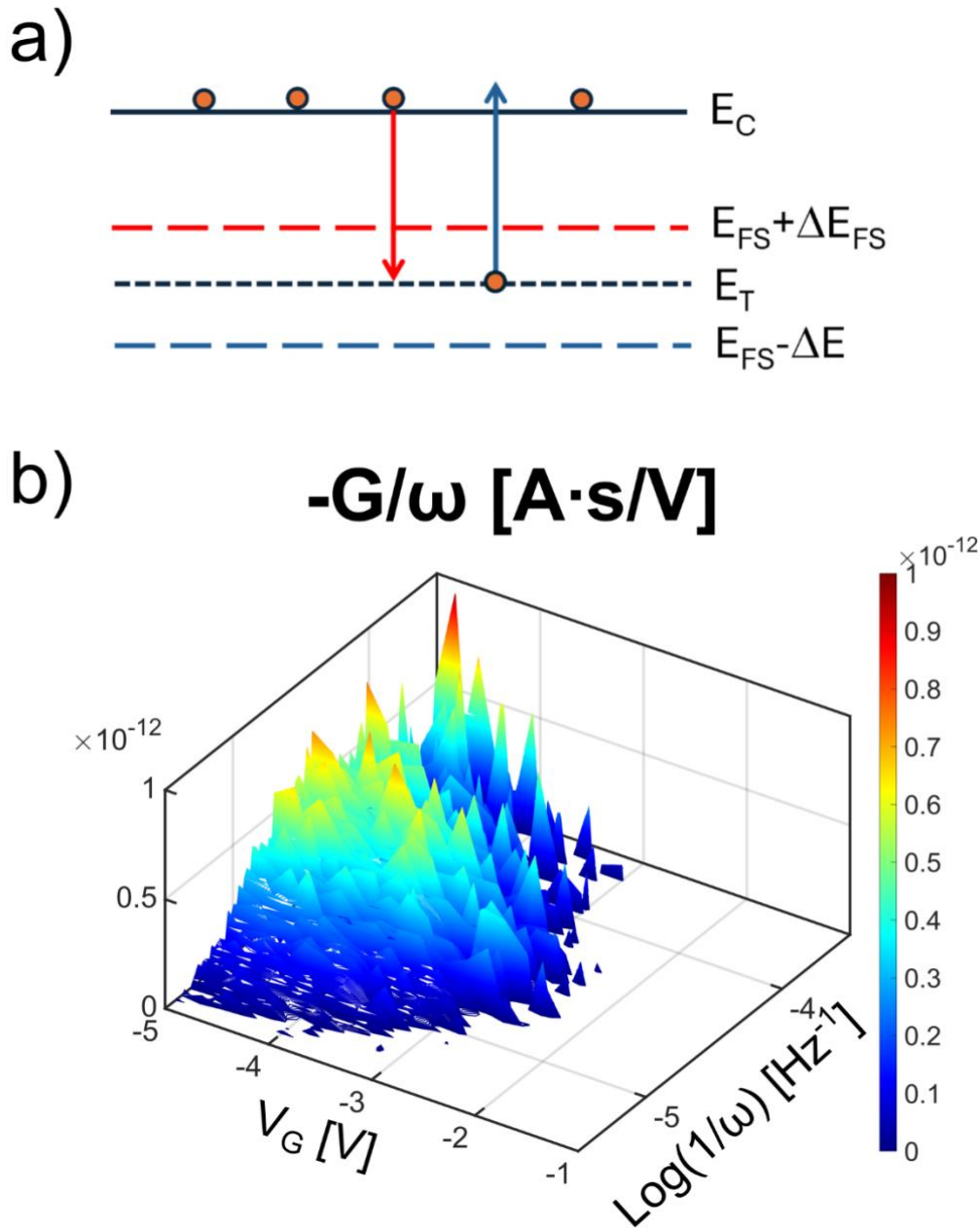

Figure 6. a) Schematic representation of the trapping mechanism observed from the experimental plot. When the quasi fermi level encounters trap levels, the injected charge gets captured in pace with the movement of the quasi fermi level, but with opposite phase with respect to the injection over the barrier. b) Negative peaks of the conductance arising from electron capture and emission, influencing the concentration of electrons in the channel with a phase angle difference of 180 degrees.